# Hot Earth or Young Venus? A nearby transiting rocky planet mystery

Kaltenegger, L.[1,2], Payne, R. C.[1,2], Lin, Z.[1,2,3], Kasting, J.[4], Delrez, L.[5]

[1] Department of Astronomy, Cornell University, 311 Space Sciences Building, Ithaca, NY 14853, USA
[2] Carl Sagan Institute, Cornell University, 302 Space Sciences Building, Ithaca, NY 14853, USA
[3] Department of Earth, Atmospheric and Planetary Sciences, MIT, Cambridge, MA 02139, USA
[4] Department of Geosciences, Penn State University, 435K Deike Building, State College, PA 16801
[5] Astrobiology Research Unit & STAR Institute, University Liege, Allee du 6 Aout 19C, B-4000 Liege, Belgium

**Abstract:** Venus and Earth provide astonishingly different views of the evolution of a rocky planet, raising the question of why these two rocky worlds evolved so differently. The recently discovered transiting Super-Earth LP 890-9c (TOI-4306c, SPECULOOS-2c) is a key to the question. It circles a nearby M6V star in 8.46 days. LP890-9c receives similar flux as modern Earth, which puts it very close to the inner edge of the Habitable Zone (HZ), where models differ strongly in their prediction of how long rocky planets can hold onto their water. We model the atmosphere of a hot LP890-9c at the inner edge of the HZ, where the planet could sustain several very different environments. The resulting transmission spectra differ considerably between a hot, wet exo-Earth, a steamy planet caught in a runaway greenhouse, and an exo-Venus. Distinguishing these scenarios from the planet's spectra will provide critical new insights into the evolution of hot terrestrial planets into exo-Venus. Our model and spectra are available online as a tool to plan observations. They show that observing LP890-9c can provide key insights into the evolution of a rocky planet at the inner edge of the HZ as well as the long-term future of Earth.

Keywords: planets and satellites: terrestrial planets, atmospheres, physical evolution, methods: numerical, observational

**Introduction:**

Venus and Earth provide astonishingly different views of the evolution of a rocky planet, raising the question of why these two rocky worlds evolved so differently. While the development of Venus and Earth are not accessible, rocky planets at the inner edge of the Habitable Zone (HZ) provide rare insights into the possible paths of climate evolution of hot rocky planets. Delrez et al. (2022) recently announced the detection of the super-Earth LP 890-9c (also SPECULOOS-2c) at the inner edge of the HZ (see e.g., Kasting 1988; Kopparapu et al. 2013, 2017; Luger &Barnes 2015; Ramirez & Kaltenegger 2014) transiting a low-activity nearby M6V star at 32 pc. The planet is part of a two-planet system with the inner planet, LP 890-9b, first detected by TESS (Guerrero et al.2021) (TOI-4306.01) in a 2.73-day orbit, with incident flux of 4.09+/-0.12 $S_\oplus$, inward of the HZ. However, LP 890-9c, receives 0.906 +/- 0.026$S_\oplus$, placing it within both the conservative and empirical HZ (Fujii et al. 2018; Kaltenegger 2017; Schwieterman et al. 2018). This makes LP 890-9c a key to understanding how Venus and Earth evolved.

To explore the range of possible atmospheres and assess whether transmission spectra can illuminate that difference, we create 7 models for LP890-9c: 2 hot exo-Earth models based on modern Earth with and without $CO_2$, 2 moist greenhouse, 1 runaway greenhouse steam, and 1 $CO_2$-dominated exo-Venus model based on modern Venus' surface pressure. We chose those models to explore the possible evolutionary progress, these individual model scenarios could represent steps in a terrestrial planet's evolution from a hot exo-Earth to an exo-Venus (Fig.1) (also Walker, 1975; Kasting, 1988; Yang et al.2013; Way et al.2016; Kopparapu et al. 2017; Way & Del Genio, 2020). In addition to the models we created, we also include a *modern Venus* atmosphere for comparison using the Venus International Reference Atmosphere (VIRA) (Moroz et al.1985; Bierson & Zhang, 2020). That *modern Venus* atmosphere has not been altered from the one in our own Solar System and contains clouds at 70km like Venus. We provide our models (Fig. A1) and spectra (Fig. 2) online as a tool to plan observations and assess the environment of this super-Earth at the inner edge of the HZ. Although the masses of LP890-9c remain to be measured ($M_P <$ 25$M_\oplus$), the discovery team estimated a mass of $M_p$ 2.5+1.8−0.8 $M_\oplus$ based on the radius-mass relationship (Chen & Kipping 2016), pointing out that LP890-9c system is the second most favorable HZ terrestrial planet system for atmospheric characterization in transmission, following the Trappist-1 system.

While there are many possible atmospheric models for exoplanets like LP890-9c (see e.g., (Fauchez et al.2021; Turbet et al.2020 for a review of models of Trappist-1), here we explore the specific question of how a rocky, Earth-like planet could evolve at LP890-9c's position and if the resulting spectra can be used to distinguish these scenarios and deepen our understanding of the conditions at the inner edge of the HZ. Our models show marked differences in the resulting transmission spectra. Observations of LP890-9c with JWST could illuminate terrestrial planet environments at the inner edge of the HZ, and also provide critical input for models to predict Earth's future.

**2. METHODS:** To assess whether different evolutionary stages of a hot rocky planet generate different transmission spectra, we model 7 scenarios for LP890-9c: Hot Earth-analogs (*Hot Earth 1&2*), moist runaway greenhouse (*Runaway 1&2*), a full runaway greenhouse/steam atmosphere (*Runaway 3*), a $CO_2$-dominated atmosphere (*$CO_2$-atm*) and we also include a modern Venus-analog (*modern Venus*). Table 1 presents relevant stellar and planetary, Table 2 atmospheric model parameters.

**Table 1: Stellar and planetary parameters**
Name: LP 890-9, TOI 4306, SPECULOOS-2
**Spectroscopic and derived properties**

| | |
|---|---|
| Optical/NIR Spectral Type: | M6.0 +/- 0.5 |
| $T_{eff}$ (K) | 2850 +/- 75 |
| $M\star$ ($M_\odot$) | 0.118 +/- 0.002 |
| $R\star$ ($R_\odot$) | 0.1556 +/- 0.0086 |
| $L\star$ ($10^{-3} L_\odot$) | 1.441 +/- 0.038 |
| Age (Gyr) | 7.2 +2.2 −3.1 |
| **Planetary parameters** | |
| Orbital Period(days) | 8.46 |
| $R_p$($R_\oplus$) | 1.367+0.055−0.039 |
| Stellar Irradiation ($S_\oplus$) | 0.906 +/- 0.026 |
| $M_P$($M_\oplus$) estimate | 2.5+1.8−0.8 |

The model scenarios are motivated by the expected stages and/or evolution of terrestrial planet at the inner edge of the HZ (Fig.1): i) hot Earth, ii) moist and full greenhouse state, iii) $CO_2$ dominated, iv) Venus (see Kasting, 1988). We first set the surface temperature ($T_S$) of the planet, assume an isothermal troposphere, and connect these to each other with a moist adiabat. Then, we do inverse calculations that change $T_S$ to achieve flux balance (see Kasting *et al.*1993). This method has been used to establish the limits of the HZ because runaway greenhouse atmospheres are highly unstable: higher $T_S$ increases water vapor concentrations, which in turn increases the $T_S$. We use this approach to simulate the evolution of $T_S$ as a function of incident irradiation, identify and model the runaway greenhouse stages. Fig. 1 shows radiation fluxes, and their derived functions versus $T_S$ for the first 5 LP890-9c models from hot Earths to greenhouse atmospheres: net flux (a) outgoing infrared ($F_{IR}$) and incident solar ($F_S$), (b) planetary albedo ($A_p$), the vertical lines indicate our models. $F_S$ and $F_{IR}$ are close for all chosen models and thus may provide temporarily stable conditions.

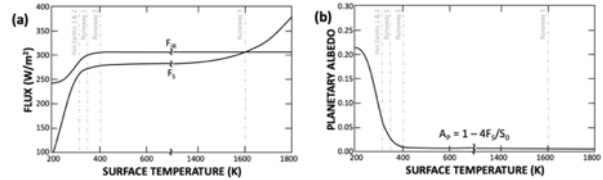

**Fig. 1:** LP890-9c model scenarios from Hot Earths to Greenhouse atmospheres: (a) Net outgoing infrared ($F_{IR}$) and incident solar flux ($F_S$), (b) planetary albedo ($A_p$) vs. surface temperature ($T_S$). Grey lines indicate the models.

The two hot Earth scenarios, "*Hot Earth 1*" and '*Hot Earth 2*', assume a modern Earth-like atmospheres (Kaltenegger *et al.*2020). We use Exo-Prime2, a well-known one-dimensional rocky exoplanet atmosphere code (Madden & Kaltenegger, 2020) that couples a climate-radiative-convective and a photochemistry model, to a line-by-line radiative transfer code (e.g. Kaltenegger *et al.*2007) to compute the temperature, atmospheric mixing ratio profiles, and spectra of all models. For the *Hot Earth 1* scenario, we set the mixing ratios of $O_2$ and $CO_2$ to 0.21 and 3.55 x $10^{-4}$ (modern Earth values), respectively, and the surface pressure to 0.97 bar, scaling the surface pressure with the planet's gravity. For the *Hot Earth 2* scenario, we reduce atmospheric $CO_2$ to negligible concentrations, assuming it has been effectively removed from the atmosphere, as expected for hot rocky planets with an active carbonate-silicate cycle. We calculate both atmospheres to 45 km height and isothermally extended beyond due to limitations of the models in the hot upper atmosphere (see Kasting *et al.* 2015).

To model moist greenhouse conditions (*Runaway 1&2*), we follow (Kopparapu *et al.* 2013) and model the atmospheres with $N_2$, $H_2O$, and $CO_2$ with the radiative-convective climate code only, where we assume that the troposphere is fully saturated with water vapor. *Runaway 1* contains 20% $CO_2$ of the dry atmosphere, a similar level to an Archean Earth. *Runaway 2* only contains modern Earth $CO_2$ levels ($10^{-6}$), assuming that $CO_2$ was effectively removed by the increased water in the atmosphere. We chose the value of 20% of the dry atmosphere mixing ratio to explore the effect of $CO_2$ on the moist greenhouse atmospheres and the spectrum on planets at the inner edge of the HZ. Note that there is no self-consistent limit on how much $CO_2$ could be in the atmosphere of a terrestrial planet in a moist greenhouse stage. If all of Earth's surface $CO_2$, including that stored in carbonate rocks, were in the atmosphere, the $CO_2$ partial pressure would be ~60 bar (Walker 1985). The *Runaway 3* scenario assumes a steam atmosphere for a planet in a full runaway greenhouse (see Ingersoll 1969; Rasool & De Bergh 1970), where all water from the oceans has evaporated into the atmosphere, resulting in a surface pressure of several hundred bars. The last two scenarios are exo-Venus models. We model a $CO_2$-dominated atmosphere, *$CO_2$-atm*, with water mixing ratio and surface pressure similar to modern Venus using inverse calculations as described above. The albedo of 0.53 for the $CO_2$ atmosphere (Fig. 1) is much higher than for water dominated/steam atmosphere (see also Kopparapu et al.2013). In addition to our models, we also include a *modern Venus* (VIRA) atmosphere for comparison, with modern Venus' atmosphere composition and known cloud height (Ignatiev et al. 2009; Pasachoff et al. 2011).

For all 7 scenarios, we generate high-resolution transmission spectra from 0.4 to 20μm at a resolution of 0.01 cm$^{-1}$ (following Kaltenegger & Traub, 2009) with Exo-Prime2. For each of the 50 layers of the model atmosphere, we calculate the line shapes and widths individually with Doppler- and pressure-broadening with several points per line width. ExoPrime2 includes Rayleigh and Mie scattering and the most spectroscopically relevant molecules from the HITRAN 2016 line list (Gordon et al. 2017). Atmospheric refraction does not restrict access to the lower atmosphere of LP 890-9c in transmission spectroscopy (see e.g. Bétrémieux &Kaltenegger, 2014, 2015; Macdonald & Cowan, 2019) but clouds that form close to or on the terminator region can obscure spectral features below the cloud layer (e.g., Kaltenegger & Traub, 2009; Robinson et al.2011). Currently, for hot Earth-like planets and runaway greenhouse scenarios, we do not have a self-consistent way to model clouds (see, e.g**.,** Jansen et al. 2019). GCM models find that clouds should concentrate on the night side for early Venus and early Earth insolation (Turbet et al. 2021), which would not influence the primary transmission spectra if they do not extend to the terminator region. While cloud research is ongoing, we only indicate the highest cloud layer on Earth at 12 km and on Venus at 70km as a dotted line in Fig.2 as a guide.

**3 RESULTS: LP890-9c Models & Spectra** Fig.A1 shows the atmospheric profiles and Fig.2 the transmission spectra of the 7 model scenarios for LP890-9c available online (zenodo): (top) hot Earths, (middle) moist and runaway greenhouse stages, (bottom) exo-Venus, summarized in Table 2.

**Table 2: Atmosphere Model Parameters**

| Model | T (K) | $P_{Surf}$ (bar) | Surface Mixing ratio $CO_2$ | $H_2O$ |
|---|---|---|---|---|
| Hot Earth 1 | 303 | 0.97 | 3.600E-04 | 1.880E-02 |
| Hot Earth 2 | 297 | 0.97 | 0.000E+00 | 1.350E-02 |
| Runaway 1 | 405 | 3.86 | 1.780E-02 | 7.410E-01 |
| Runaway 2 | 340 | 1.27 | 2.950E-04 | 2.140E-01 |
| Runaway 3 | 1600 | 271.00 | 1.200E-06 | 9.960E-01 |
| $CO_2$-atm | 593 | 93.00 | 9.999E-01 | 9.877E-06 |
| Venus | 735 | 92.10 | 9.650E-01 | 1.000E-05 |

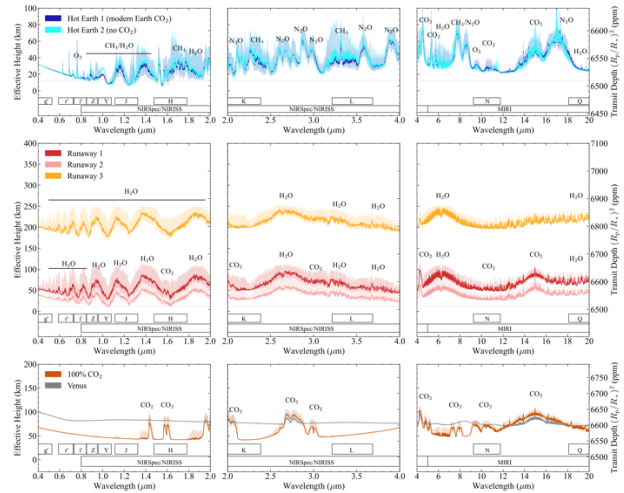

**Fig. 2:** Model transmission spectra of LP890-9c: (top) Hot exo-Earths, (middle) moist and full runaway greenhouse, and (bottom) exo-Venus models ($CO_2$-atm and modern Venus) shown at a resolution of 1000 (solid) and 10,000 (light). The highest Earth clouds at 12 km and Venus clouds at 70km are indicated as a dashed line.

The 2 *Hot Earth* scenarios (top) use modern Earth models for modern and reduced $CO_2$ (*Hot Earth 1 & 2, respectively*) at the orbital distance of LP890-9c. Lower atmospheric $CO_2$ results in slightly lower $T_S$. Major absorption features of $H_2O$, $CO_2$, $O_2$, $O_3$, $CH_4$, and $N_2O$ are labeled. Features of $N_2O$, $CH_4$, and $H_2O$ are distributed throughout the spectrum. For $O_2$ the strongest feature is found in the visible, for $O_3$ and $CO_2$ in the IR. Assuming an effective removal of $CO_2$ in the *"Hot Earth 2"* scenario $CO_2$ absorption features disappear from the transmission spectrum. Note that the $N_2O$ absorption feature is among the most dominant features in the Hot Earths spectra (see also Segura et al. 2005). The surface flux of $N_2O$ was fixed in the model at modern Earth value; but the minimal near-UV radiation from an inactive M star effectively slows $N_2O$ photolysis enough that it can accumulate. The buildup of $N_2O$ only has a minor impact on $T_S$, especially when compared to $H_2O$ and $CO_2$ in a Hot Earth scenario. The two moist runaway greenhouse scenarios, *Runaway 1&2* (middle), show $H_2O$ vapor increase with increasing $T_S$, which in turn increases the total surface pressure ($P_{Surf}$) as expected for terrestrial planet atmospheres near the inner edge of the HZ: leading to a $P_{Surf}$ of 1.27 bar and 3.46 bar for *Runaway 2* (20% $CO_2$ of the dry atmosphere) and *Runaway 1* (modern Earth $CO_2$), respectively. The stratospheric temperature warms with increasing $CO_2$, and stratospheric water vapor increases with $T_S$ (see Leconte *et al.*2013; Kasting *et al.* 2015; Wolf & Toon, 2015). *Runaway 3* shows a full runaway greenhouse steam atmosphere, with a $P_{Surf}$ of 271bar, and $T_S$ of 1600K.

The transmission spectra (Fig.2) show strong $H_2O$ features throughout the wavelength range modeled, with the strongest $CO_2$ feature in the IR. Increasing water vapor concentration in the Runaway atmospheres increases the transit depth and the effective height of the individual absorption features, making the full runaway greenhouse model, *Runaway 3*, the atmosphere with the largest effective height and transit depth. Note that there is no self-consistent modeling of how much $CO_2$ should accumulate in the atmosphere of a hot terrestrial planet. Thus, detecting the $CO_2$ features in combination with the $H_2O$ features can address the question of whether $CO_2$ will be effectively removed during the runaway greenhouse stage. The two $CO_2$-dominated atmosphere models (bottom) show i) a $CO_2$-atmosphere (*$CO_2$-atm*) based on modern Venus (modeled with $N_2$, $H_2O$, and $CO_2$) that produces a $T_S$ of 599K for a $P_{Surf}$ of 93bar, like on modern Venus. We also added ii) the VIRA Venus atmosphere for observed Venus atmospheric conditions (*modern Venus*) placed on LP890-9c for comparison with 92bar $P_{Surf}$ and 735K $T_S$. The biggest difference in the spectra of the $CO_2$ atmospheres is due to the known cloud-layer at 70km for modern Venus, which strongly affects the depth of the spectral features. The *modern Venus* spectrum is dominated by clouds as well as Mie scattering, which flattens out the spectra compared to the *$CO_2$-atm*, where we did not include clouds or Mie scattering due to the unknown height of any clouds that could develop. The effective height of the planetary atmosphere ranges up to 80 km for the hot Earths, up to 120 km for the moist greenhouse, up to 250 km for the runaway greenhouse, and up to 100 km for the $CO_2$-dominated Venus-like atmospheres spectra. Note that the modern Venus atmosphere shows only a maximum effective height of the absorption features of about 20km due to the high cloud deck, with points out how severely modern Venus-like cloud decks will limit any characterization of modern Venus-analogs, compared to atmospheres without cloud coverage at the terminator region. The transit depth of the LP890-9c models are about 6,600 ppm; the largest absorption feature depths are about 100 ppm deep. The models with a higher effective height show bigger transit depth $(R_p/R_s)^2$.

**4. Discussion:** How a hot Earth-like planet could evolve into a modern Venus is still an area of active research with many open questions on critical issues, such as how fast such a process could happen and what atmospheric compositions and cloud properties of such evolutionary stages would be. Our models show a variety of possible atmospheres on such a path. If time evolution were to be included, assuming an Earth-like water ocean, most of the water-rich atmospheres would likely have lost their water due to hydrodynamic escape by this late in the parent star's history, based on current age estimates for LP890-9c. For an Earth-analog planet, we would expect it to either remain in a hot Earth stage, where the planet would not lose its water to space or to have already transitioned to a Venus-like state. Thus, we included a modern Venus model in our comparison as well. But the initial water inventory of LP890-9c is unknown. If it started its life as a mini-Neptune, then water and other volatiles might still be present despite large losses over time. We also do not know how much EUV energy is, or was, available to drive such an escape. So, we could only speculate on the timescale of water. Thus, we have also included models for greenhouse stages because the evolution through these stages has not yet been observed.

Recently, 3D global climate model (Turbet et al. 2021) simulated beyond what solar irradiation surface water can no longer condense on a Young Venus and a Young Earth (see also Hamano et al. 2015; Kasting et al. 2014; Leconte et al. 2013; Wolf & Toon 2015). Around a star like the Sun, the planet develops (or remains in) a runaway greenhouse at incident stellar irradiation larger than 0.95 times modern Earth's irradiation ($S_\oplus$) for a Young Venus (modeled as a slow rotator of 5833 hours) and 0.92 $S_\oplus$ for a Young Earth (modeled as a faster rotator of 24 hours). This means stellar irradiation greater than that should have kept water from condensing on a young rocky world, prohibiting oceans to form and initiating catastrophic water loss. Intriguingly, the incident irradiation at LP890-9c's position, 0.906 $S_\oplus$, is below but close to these limits. However, note that the irradiation of a cooler red star is more effective in heating the surface of a planet than the irradiation by a sun-analog star (Kasting et al.1993). Thus, the limits of "no water condensation" should appear at lower incident irradiation for cooler stars.

In general, surface UV levels on M star planets should span a wide range (e.g., Scalo *et al.*, 2007; Tarter *et al.*, 2007; Shields et al. 2016; O'Malley-James & Kaltenegger, 2019), which could require life to either shelter or evolve mitigation strategies for high UV levels (e.g., Luger & Barnes 2015; O'Malley-James & Kaltenegger 2018, 2019b; Ramirez & Kaltenegger 2014) influencing its remote detectability.

**5. Conclusions:** The recently discovered transiting Super-Earth LP 890-9c is a key to the question of how Venus and Earth developed so vastly differently. LP890-9c receives about 0.9 times the flux of modern Earth, putting it very close to the inner edge of the Habitable Zone, where models differ strongly in their prediction of how long Earth-like planets can hold on to their water and what environment such planets could sustain. To address this, we modeled 7 scenarios, covering the possible evolutionary stages of rocky planets at the inner edge of the HZ for LP890-9c: Our models show marked differences in the climate and resulting transmission spectra of chemicals that can reveal the difference between hot exo-Earths, planets in runaway greenhouse stages, a CO2-dominated, and a modern exo-Venus atmosphere. Distinguishing these scenarios from the planet's spectra can provide critical new insights into how fast terrestrial planets lose their water, the evolution of rocky planets at the inner edge of the HZ, and the future evolution of our own planet. LP890-9c provides a rare opportunity to explore the evolution of terrestrial planets at the inner edge of the HZ and is a prime target for observations with telescopes like JWST.

**Acknowledgements:** LK & RP thank the Brinson Foundation. L.D. F.R.S.-FNRS Postdoctoral Researcher.

**Data Availability:** The data underlying this article are available (insert zenodo link here when accepted)